\documentclass[english,a4paper,12pt]{article}

\usepackage{threeparttable}
\usepackage[margin=24mm]{geometry}
\usepackage{amsmath,amssymb}
\IfFileExists{algorithm.sty}{\usepackage{algorithm}}{}
\IfFileExists{algorithmic.sty}{\usepackage{algorithmic}}{}
\usepackage{enumitem}
\usepackage[title]{appendix}
\usepackage{array}
\usepackage{graphicx}
\usepackage{tikz}
\usetikzlibrary{arrows.meta,positioning,calc}
\usepackage{amsxtra, amsfonts, amssymb, amstext, amsmath, mathtools}
\usepackage{amsthm}
\usepackage{booktabs}
\usepackage{nicefrac}
\usepackage{xspace}
\usepackage{url}
\usepackage{graphics,xcolor}
\IfFileExists{algorithm2e.sty}{\usepackage[algo2e,ruled,vlined,linesnumbered]{algorithm2e}}{}
\usepackage{wrapfig}
\IfFileExists{bbm.sty}{\usepackage{bbm}}{}
\usepackage{mathtools}
\usepackage[hypertexnames=false]{hyperref}
\usepackage{longtable}
\usepackage{fancyhdr}
\usepackage{needspace}
\usepackage{listings}
\usepackage{float}

\newtheorem{theorem}{Theorem}[section]
\newtheorem{lemma}[theorem]{Lemma}
\newtheorem{proposition}[theorem]{Proposition}
\newtheorem{corollary}[theorem]{Corollary}

\newtheorem{remark}[theorem]{Remark}
\newtheorem{example}[theorem]{Example}
\theoremstyle{definition}
\numberwithin{equation}{section}

\let\originalleft\left
\let\originalright\right
\renewcommand{\left}{\mathopen{}\mathclose\bgroup\originalleft}
\renewcommand{\right}{\aftergroup\egroup\originalright}

\definecolor{linkblue}{RGB}{24,57,93}
\hypersetup{colorlinks=true,linkcolor=linkblue,citecolor=linkblue,urlcolor=linkblue,
 pdftitle={Exact Quantum Query Complexity of the Five-Bit Symmetric Boolean Function EXACT(5;3,4)}}
\DeclareMathOperator{\Tr}{Tr}
\DeclareMathOperator{\Exact}{EXACT}
\newcommand{\QE}{Q_{\mathrm E}}
\newcommand{\sosdeg}{\deg_{\mathrm{sos}}}
\newcommand{\one}{\boldsymbol{1}}
\newcommand{\cS}{\mathcal S}
\newcommand{\Spalek}{\v{S}palek}
\newcommand{\Uyanik}{Uyan{\i}k}
\allowdisplaybreaks[1]
\title{Exact Quantum Query Complexity of the Five-Bit Symmetric Boolean Function $\Exact^5_{3,4}$\thanks{Corresponding authors: Guoliang Xu (xu1guo2liang@foxmail.com).}}
\author{Guoliang Xu, Haoran Zhang, Mengfan Wu, Songyao Xue\\ School of Software\\ Luoyang Normal University\\ Luoyang 471934, China
\and Mingke Liu\\ School of Artificial Intelligence\\ Luoyang Normal University\\ Luoyang 471934, China
}

\begin{document}
\maketitle
\begin{abstract}
Let \(f = \text{EXACT}^5_{3,4}\) be the Boolean function that equals 1 exactly on inputs of Hamming weight 3 or 4. Earlier numerical studies reported success probabilities close to one using three queries, but the existence of an exact three-query algorithm had remained unresolved. This paper settles this problem by proving that the exact quantum query complexity of \(f\) is \(4\).
For the lower bound, we express every three-query algorithm through layer-by-layer Gram matrices of low-degree Fourier coefficients. We then construct a fixed rational dual certificate whose slack matrices are all positive definite. This certificate separates the necessary Gram-matrix region from the exact-output condition and therefore rules out every exact algorithm using at most three queries in the stated model. The positivity verification reduces to five small rational matrices and Sylvester's criterion. For the upper bound, we invoke the previously known four-query algorithm for the seven-bit function \(\text{EXACT}^7_{3,4}\) and apply the standard padding reduction.
To quantify the near-exact behavior, we use the same certificate to derive a valid,
though not claimed tight, lower bound
\(\frac{25600}{43779488192449}>0\)
on the worst-case error of every algorithm using at most three queries. This positive error gap is compatible with the near-unit success
probabilities reported in earlier numerical studies, which do not
establish exact computation.
\end{abstract}

\section{Introduction}
The exact quantum query complexity \(Q_{\mathrm E}(f)\) measures the minimum number of oracle calls needed to compute a Boolean function \(f\) correctly with probability one on every input. It differs from the usual bounded-error query complexity, which allows a certain probability of error. Exact complexity is usually harder to determine because numerical methods, even when they give highly accurate results, cannot directly prove exactness.

For small functions, numerical methods such as semidefinite programming can find highly accurate algorithms, but rounding a numerical output to \(1\) does not constitute a rigorous proof. In this paper, we resolve the remaining three-versus-four gap by combining the previously known four-query upper bound with a new impossibility certificate for every algorithm using at most three queries in the stated model.

We study the function
\begin{equation}\label{eq:function}
 f(x)=\Exact^5_{3,4}(x)=
 \begin{cases}1,&|x|\in\{3,4\},\\0,&|x|\in\{0,1,2,5\},\end{cases}
 \qquad x\in\{0,1\}^5,
\end{equation}
where \(|x|=\sum_{j=1}^5x_j\).  Methodologically, we rely on the polynomial representation of Beals et al. The amplitudes of a \(T\)-query algorithm are multilinear polynomials of degree at most \(T\), and hence its acceptance probability has degree at most \(2T\) \cite{BBCMW}. Nisan and Szegedy \cite{NisanSzegedy} established foundational relations between polynomial degree and classical decision-tree complexity. Gram-matrix and semidefinite-programming formulations of query algorithms were developed by Barnum, Saks, and Szegedy \cite{BSS}, and are closely related to adversary methods \cite{Ambainis03,Reichardt11,LeeMittalReichardtSpalekSzegedy11}. \Spalek{} and Szegedy \cite{SpalekSzegedy} proved the
equivalence of the then-known adversary formulations. These tools provide the foundation for the exactness arguments in this paper.

Numerical studies of small exact-query problems include Montanaro, Jozsa, and Mitchison \cite{MJM} and \Uyanik{}~\cite{Uyanik}. The work of Montanaro, Jozsa, and Mitchison first appeared
as an arXiv preprint in 2011; its journal version appeared
in Volume~71 of \emph{Algorithmica} in 2015~\cite{MJM}.
Appendix~A.2 of the 2011 preprint lists the five-bit
symmetric function with weight-value vector
$(0,0,0,1,1,0)$, indexed by Hamming weights
$0,1,\ldots,5$.
This is exactly the function studied here.
Its three-query entry is $1$; under the authors' stated
convention, this denotes a solver-reported success
probability greater than $0.999$, rather than an exact
value of one~\cite[arXiv v1, Appendix~A]{MJM}. This numerical evidence therefore neither proves nor disproves the existence of a three-query exact algorithm. \Uyanik{} later studied exact query complexity using a generalized SDP approach~\cite{Uyanik}. Ambainis, Iraids, and Nagaj \cite[Theorem~1]{AIN} proved that $\QE(\Exact^{2k+1}_{k,k+1})=k+1$ for $k\ge1$. Taking $k=3$ gives a four-query algorithm for $\Exact^7_{3,4}$. Since $\Exact^5_{3,4}(x)=\Exact^7_{3,4}(x_1,\ldots,x_5,0,0)$, padding with two fixed zeros yields $\QE(\Exact^5_{3,4})\le4$. Combined with the elementary lower bound, the exact complexity of the present function is therefore either three or four.

He, Sun, Yang, and Yuan studied the related partial weight-decision problem using Chebyshev polynomials and a quantum-padding technique, but their promise-function results are distinct from the present total-function certificate \cite{HeSunYangYuan}. Xu and Qiu \cite{XQ} studied the relation between SOS representations and exact quantum query algorithms. For the present function, they showed that two specific pairs of SOS representations of $f$ and $1-f$ cannot produce an exact three-query algorithm through their construction. This excludes those representations within that construction, but does not determine $\QE(f)$. Yao and Ye \cite{YY} obtained tight results for additional subfamilies of $\Exact^n_{k,l}$, where the function equals $1$ exactly when $|x|\in\{k,l\}$. Their new EXACT results include the cases $l-k\ge4$ with $k=0$, and $l-k\ge4$ with $k=1$ and $l=n-1$. These cases do not include the adjacent-weight instance $(n,k,l)=(5,3,4)$ considered here. Podder, Yao, and Ye studied fine-grained quantum and randomized query complexity for symmetric functions, including success probabilities close to $1/2$; that broader setting provides context for, but does not replace, the present exact certificate \cite{PodderYaoYe}. 

Structural results for functions with low exact quantum query complexity provide another relevant line of work. Xu and Qiu \cite{XuQiuEntropy2021} gave necessary and sufficient conditions for partial Boolean functions to have exact quantum query complexity one. Qiu and Xu \cite{QiuXuSpin2021} reviewed exact one-query algorithms and constructed a partial Boolean function that can be computed with one quantum query but not by the Deutsch--Jozsa algorithm. Xu and Qiu \cite{XuQiuPolynomial2026} developed polynomial
representations under an equivalence relation for such functions
and analyzed instances on up to four input bits.
For two queries, Ye, Yang, and Huang \cite{YeYangHuang} studied
the classification of Boolean functions with exact quantum query
complexity two. Their classification relies on the correctness
of the numerical classification for four-bit functions.

Earlier exact results for EXACT and THRESHOLD functions are due to Ambainis, Iraids, and Smotrovs \cite{AmbainisIraidsSmotrovs}; Midrijanis studied general total Boolean functions \cite{Midrijanis04}. For promise functions defined by weight layers, Qiu and Zheng
\cite{QiuZheng} studied a generalization of the Deutsch--Jozsa
problem and its optimal quantum algorithm.
Xu et al. \cite{XuEtAlDJ2024} studied the classical randomized
decision tree complexity of the Deutsch--Jozsa problem and a
generalization, highlighting the role of allowed error in
comparisons between classical and quantum query complexity. 

Wu, Hou, Zhang, Li, and Zeng \cite{WuHouZhangLiZeng}
developed variational learning methods for quantum query algorithms and numerically investigated additional EXACT instances, including noncentral adjacent-weight cases. These numerical studies provide further motivation for exact certificates that distinguish near-unit success probabilities from exact computation. Taken together, these results motivate the universal three-query certificate developed below.

The main difficulty is to rule out every three-query process, rather than only a particular circuit ansatz. Equivalently, proving \(\QE(\Exact^5_{3,4})=4\) amounts to excluding every exact algorithm using at most three queries in the stated model. We therefore retain pairwise final-state inner products in a low-degree Fourier basis and use a fixed rational certificate to separate the resulting necessary region from exact-output Gram matrices. The certificate applies to arbitrary finite-dimensional ancillas, arbitrary complex amplitudes, and arbitrary input-independent unitaries; symmetry is used only to compress the certificate and its verification.

The main theorem resolves this three-versus-four gap.
\begin{theorem}\label{thm:main}
For the function $f$ in \eqref{eq:function}, the exact quantum query complexity in the standard model is
\begin{equation}\label{eq:intro-complexity}
\QE(f)=4.
\end{equation}
\end{theorem}

The new ingredient is an explicit rational dual certificate excluding every exact three-query algorithm. The four-query upper bound is inherited from prior work.

The remainder of the paper is organized as follows. Section~\ref{sec:gram} introduces the phase-query model, Fourier notation, the necessary Gram recurrence, and the dual infeasibility lemma. Section~\ref{sec:certificate} gives the explicit rational certificate, compresses its verification using permutation orbits, proves positivity of all slack matrices, and combines these ingredients to complete the proof of the main theorem. Section~\ref{sec:consequences} records the symmetric degree-three SOS representation and derives the quantitative lower bound on the error of every algorithm using at most three queries. Section~\ref{sec:conclusion} concludes the paper. Appendix~\ref{app:background} collects the standard background and notation used throughout. The remaining appendices contain the expanded Fourier--Gram and telescoping proofs, the block-matrix positivity calculations, and the exact verification protocol, followed by the complete embedded source code.

\section{Preliminaries and the Fourier--Gram Framework}\label{sec:gram}
This section develops the algebraic framework used in the lower-bound proof. We first pass to an equivalent phase-query model and introduce the low-degree Fourier basis. We then derive the Gram-matrix recurrence and formulate the dual infeasibility lemma used in Section~\ref{sec:certificate}. Readers unfamiliar with these terms may consult Appendix~\ref{app:background}, which gives short definitions and proofs of the standard facts used below.

\subsection{Phase queries and Fourier notation}
Let $X=\{0,1\}^5$ and write $[5]=\{1,2,3,4,5\}$. The standard bit query is
\begin{equation}\label{eq:bit-query}
 O_x^{\mathrm{bit}}|i,b,w\rangle=|i,b\oplus x_i,w\rangle,\qquad 1\le i\le5.
\end{equation}
Here $i$ is the query-index register, $b$ is the answer bit, and $w$ denotes the workspace; $\oplus$ denotes XOR, i.e. addition modulo $2$.
In the Hadamard basis of the answer qubit, $|+\rangle$ is unchanged and $|-\rangle$ acquires the phase $(-1)^{x_i}$. We therefore use the equivalent phase-query model with a virtual index $0$ and $x_0=0$:
\begin{equation}\label{eq:phase-query}
 O_x|i,w\rangle=(-1)^{x_i}|i,w\rangle,\qquad 0\le i\le5.
\end{equation}
The conversion is coherent and preserves the original registers in the workspace. To simulate a bit query, compute a fresh phase-query index: use index $0$ on the $|+\rangle$ answer branch and index $i$ on the $|-\rangle$ branch. Apply the phase query and uncompute the fresh index. Conversely, retain the original phase-query index in the workspace. For $i\ge1$, prepare a fresh bit-query register with index $i$ and answer state $|-\rangle$; for $i=0$, use the fixed legal index $1$ and answer state $|+\rangle$.
Apply one bit query and uncompute the auxiliary registers. These input-independent reversible operations implement both simulations with one oracle call, so the two models have the same query complexity. All other operations are arbitrary input-independent unitaries on an arbitrary finite-dimensional workspace. Intermediate measurements and internal randomness may be purified and stored coherently. This state-decomposition viewpoint is closely related to the output-Gram description studied by Grillo and Marquezino \cite{GrilloMarquezino}. The standard query-model reductions used in this paragraph are summarized in Appendix~\ref{app:query-model}.

For $S\subseteq[5]$, let
\begin{equation}\label{eq:character}
 \chi_S(x)=(-1)^{\sum_{j\in S}x_j}
\end{equation}
be the Fourier character, and set
\begin{equation}\label{eq:subset-family}
 \cS_t=\{S\subseteq[5]:|S|\le t\},\qquad d_t=|\cS_t|,
 \qquad (d_0,d_1,d_2)=(1,6,16).
\end{equation}
Let $F_t\in\mathbb R^{32\times d_t}$ be $(F_t)_{x,S}=\chi_S(x)$. Character orthogonality gives
\begin{equation}\label{eq:character-orthogonality}
 F_t^{\mathsf T}F_t=32I_{d_t}.
\end{equation}
We use $[1]$ and $[-1]$ for $1\times1$ matrices.
Put $q_0=\varnothing$, $q_i=\{i\}$ for $i\ge1$, and let $D_i:=\operatorname{diag}(\chi_{q_i}(x))_{x\in X}$. Throughout this section, an unqualified sum over $i$ means $i=0,1,\ldots,5$. We write $I_d$ for the $d\times d$ identity matrix and $I$ for the identity operator. Let $\one_E$ denote the indicator of a condition $E$ (equal to $1$ when $E$ holds and $0$ otherwise). For $t=0,1$, define
\begin{equation}\label{eq:T}
 (T_{t,i})_{R,S}=\one_{\{R=S\triangle q_i\}},
 \qquad R\in\cS_{t+1},\ S\in\cS_t.
\end{equation}
Thus $T_{t,i}$ maps a coefficient vector of degree at most $t$ to one of degree at most $t+1$. Then, for $t=0,1$,
\begin{equation}\label{eq:Fourier-query-action}
 F_{t+1}T_{t,i}=D_iF_t,
\end{equation}
because $\chi_{q_i}\chi_S=\chi_{S\triangle q_i}$. The matrices $T_{t,i}$ describe the query in coefficient space, while $D_i$ describes the same operation on value vectors. We write $F=F_2$ below.

The ordering of $X$ and of each subset family is arbitrary
but fixed. Changing these orderings permutes the corresponding
rows and columns. For square matrices whose rows and columns
are indexed by the same family, this is conjugation by a
permutation matrix. The rectangular matrices $F_t$ and
$T_{t,i}$ instead undergo the corresponding row and column
permutations. All identities and positivity statements are
invariant under these changes. The characters satisfy
\begin{equation}\label{eq:character-product}
 \chi_S(x)\chi_R(x)=\chi_{S\triangle R}(x),
 \qquad \chi_S(x)^2=1,
\end{equation}
where $S\triangle R$ is symmetric difference. For $S\ne R$, choose an element of $S\triangle R$ and pair inputs differing in that coordinate; the corresponding character product changes sign. This proves the orthogonality relation and explains why $F_t/\sqrt{32}$ has orthonormal columns. The resulting Fourier identities are collected in Appendix~\ref{app:fourier} for reference.

The definition of $T_{t,i}$ is well posed because $|S\triangle q_i|\le |S|+1\le t+1$. For a coefficient vector $a$,
\begin{equation}\label{eq:query-action-expanded}
\begin{aligned}
 (F_{t+1}T_{t,i}a)_x
 &=\sum_{S\in\cS_t}a_S\chi_{S\triangle q_i}(x)\\
 &=\chi_{q_i}(x)\sum_{S\in\cS_t}a_S\chi_S(x)
 =(D_iF_ta)_x.
\end{aligned}
\end{equation}
Thus $T_{t,i}$ and $D_i$ are the coefficient-space and value-space forms of the same query action.

\subsection{A necessary condition for three queries}
With this notation in place, we now translate an arbitrary three-query computation into necessary algebraic constraints. The Gram-matrix facts used in the next lemma are reviewed in Appendix~\ref{app:gram}. The measurement and output-orthogonality fact used in the exact-output clause is summarized in Appendix~\ref{app:povm}.

\begin{lemma}[Gram recurrence]\label{lem:necessary}
Let $f:\{0,1\}^5\to\{0,1\}$. Every algorithm making at most three queries gives complex Hermitian positive semidefinite (PSD) matrices
\begin{equation}\label{eq:gram-dimensions}
 A_{t,i}\in\mathbb C^{d_t\times d_t},\qquad t=0,1,2,\quad i=0,1,\ldots,5,
\end{equation}
satisfying
\begin{align}
 \sum_iA_{0,i}&=[1],\label{eq:init}\\
 \sum_iA_{t+1,i}&=\sum_iT_{t,i}A_{t,i}T_{t,i}^{\mathsf T},\qquad t=0,1,\label{eq:recurrence}\\
 G&=\sum_iD_iFA_{2,i}F^{\mathsf T}D_i.\label{eq:G}
\end{align}
Here $G_{xy}=\langle\psi_x|\psi_y\rangle$ is the Gram matrix of the purified final states. If the algorithm computes $f$ exactly, then $G_{xy}=0$ whenever $f(x)\ne f(y)$.
\end{lemma}

\begin{proof}
Purify the initial state, internal randomness, and intermediate measurements. If a branch stops early, add a virtual index-$0$ query, which acts as the identity and therefore does not change the computation. Thus the computation has three phase-query slots.

Let $|\phi_t(x)\rangle=\sum_i|i\rangle|a_{t,i}(x)\rangle$ be the state before the $(t+1)$st query. By induction there are input-independent vectors $|v_{t,i,S}\rangle$ such that
\begin{equation}\label{eq:state-expansion}
 |a_{t,i}(x)\rangle=\sum_{S\in\cS_t}\chi_S(x)|v_{t,i,S}\rangle.
\end{equation}
The base case is the input-independent initial state. A query multiplies a branch by $\chi_{q_i}$ and sends $S$ to $S\triangle q_i$, increasing the degree by at most one; an input-independent unitary only takes linear combinations of coefficient vectors.

Define
\begin{equation}\label{eq:gram-entry}
 (A_{t,i})_{S,R}=\langle v_{t,i,S}\mid v_{t,i,R}\rangle.
\end{equation}
These matrices are Gram matrices, hence positive semidefinite even when the vectors and amplitudes are complex. Initial normalization gives \eqref{eq:init}. Equality of the input-state inner products immediately before and after an input-independent unitary, together with \eqref{eq:Fourier-query-action} and $F_{t+1}^{\mathsf T}F_{t+1}=32I$, gives \eqref{eq:recurrence}. The final query gives \eqref{eq:G}.

For exact computation, let $E_0,E_1$ be the effects of a positive-operator-valued measure (POVM), with $E_0+E_1=I$, and let $E_0$ correspond to output $0$. If $f(x)=0$, then $\langle\psi_x|E_0|\psi_x\rangle=1$, so $(I-E_0)|\psi_x\rangle=0$; if $f(y)=1$, then $E_0|\psi_y\rangle=0$. Consequently $G_{xy}=\langle\psi_x|E_0|\psi_y\rangle=0$. The lemma is a necessary condition; no converse is needed in the lower-bound proof.
\end{proof}

The full seven-step derivation is collected in Appendix~\ref{app:proof-details}.

\begin{remark}
The matrices $A_{t,i}$ retain all pairwise input inner products through the coefficient Gram structure. Looking only at diagonal acceptance probabilities would lose the recurrence. The dimensions $1,6,16$ depend on the Fourier degree, not on the ancilla dimension.
\end{remark}

\subsection{A dual infeasibility certificate}
The Gram recurrence describes the semidefinite feasibility region forced by any three-query computation. We next construct a dual certificate that rules out this entire region under the exact-output condition. The separating-functional interpretation of this certificate is explained in Appendix~\ref{app:dual}.

\begin{lemma}[Dual certificate]\label{lem:dual}
Let $H_t\in\mathbb R^{d_t\times d_t}$, $t=0,1,2$, and $C\in\mathbb R^{32\times32}$ be real symmetric matrices, and take $H_0=[-1]$. Assume $C_{xy}=0$ whenever $f(x)=f(y)$. Define
\begin{align}
 Z_{t,i}&=H_t-T_{t,i}^{\mathsf T}H_{t+1}T_{t,i},&&t=0,1,\label{eq:Z01}\\
 Z_{2,i}&=H_2+F^{\mathsf T}D_iCD_iF.&&\label{eq:Z2}
\end{align}
If the six matrices $Z_{t,i}$ at each layer $t=0,1,2$ are positive semidefinite, then no exact algorithm making at most three queries can compute $f$.
\end{lemma}
\begin{proof}
Assume a three-query exact algorithm and its matrices $A_{t,i}$. Because both $Z_{t,i}$ and $A_{t,i}$ are PSD, $\Tr(Z_{t,i}A_{t,i})\ge0$; this remains true for complex Hermitian $A_{t,i}$ because
$\Tr(ZA)=\Tr(Z^{1/2}AZ^{1/2})\ge0$.
Summing over $t,i$ and using the recurrence telescopes all $H_1,H_2$ terms. The terms involving $H_1$ and $H_2$ cancel by \eqref{eq:recurrence}, leaving only the initial condition and the final Gram term:
\begin{align}
\sum_{t=0}^2\sum_i\Tr(Z_{t,i}A_{t,i})
={}&\Tr\left(H_0\sum_iA_{0,i}\right)\\
&+\sum_{t=1}^2\Tr\left[H_t\left(\sum_iA_{t,i}-\sum_iT_{t-1,i}A_{t-1,i}T_{t-1,i}^{\mathsf T}\right)\right]
+\Tr(CG)\nonumber\\
={}&-1+\Tr(CG).\label{eq:telescoping}
\end{align}
The trace is $\Tr(CG)=\sum_{x,y}C_{xy}G_{yx}$.
If $f(x)=f(y)$, then $C_{xy}=0$; otherwise exactness gives
$G_{yx}=0$. Hence $\Tr(CG)=0$, while the left side is
nonnegative, a contradiction.
\end{proof}

Arunachalam, Briët, and Palazuelos \cite{ABP} characterize quantum query algorithms through completely bounded forms. The proof above instead proceeds directly from the necessary Fourier--Gram constraints and the displayed telescoping identity. The expanded trace calculation is collected in Appendix~\ref{app:proof-details}.

Thus the lower-bound problem has been reduced to constructing one certificate and verifying its positive semidefiniteness. Section~\ref{sec:certificate} gives this certificate explicitly and proves the required positivity statements.

\section{The Rational Certificate and the Main Theorem}\label{sec:certificate}

This section constructs the dual certificate required by Lemma~\ref{lem:dual} and completes the proof of the main theorem. The certificate consists of real symmetric matrices $H_0,H_1,H_2,C$, specified by two rational orbit tables. We first express the lower-bound problem as a finite-dimensional dual functional and define the slack matrices. We then give the certificate data, use $S_5$ symmetry to compress the eighteen slack matrices to six representatives, compute the required Fourier entries exactly, and verify positive definiteness via Sylvester's criterion. To conclude, we combine the resulting lower bound with the known four-query upper bound to prove $\QE(f)=4$.

\subsection{Certificate as a finite-dimensional dual witness}
The necessary conditions in Lemma~\ref{lem:necessary} define a finite-dimensional semidefinite feasibility region. If an exact algorithm exists, it produces a point in this region. The lower-bound proof uses only this implication and makes no converse claim. Fix $H_0=[-1]$ and real symmetric matrices $H_1,H_2,C$, and define the slacks as in Lemma~\ref{lem:dual}. Set
\begin{align}
 \mathcal L(\{A_{t,i}\}):={}&\sum_{t=0}^{2}\sum_{i=0}^{5}\operatorname{Tr}(Z_{t,i}A_{t,i})\\
 ={}&-1+\operatorname{Tr}(CG).
 \label{eq:dual-functional}
\end{align}
The second equality follows from the initialization $\sum_i A_{0,i}=[1]$, the Gram recurrence, and the terminal Gram identity in Lemma~\ref{lem:necessary}. If $C$ vanishes on equal-label pairs, exactness gives $\mathcal L(\{A_{t,i}\})=-1$, whereas PSD slacks give $\mathcal L(\{A_{t,i}\})\ge0$.

One may interpret $H_t$ and $C$ as variables of an SDP dual, but the proof does not use strong duality or strict feasibility of an abstract program. It directly verifies a fixed collection of rational matrices and exact inequalities. Numerical optimization can be used to discover a candidate, but the theorem depends only on the recovered rational data. This distinction prevents numerical near-feasibility from being confused with an exact algorithm.

\subsection{Certificate data and orbit tables}
We now specify the certificate data. Take $H_0=[-1]$. For $S,R\in\cS_t$, $t=1,2$, set
\begin{equation}\label{eq:subset-orbit-parameters}
 p=\min(|S|,|R|),\quad q=\max(|S|,|R|),\quad r=|S\cap R|,
\end{equation}
and define $(H_t)_{S,R}=h_t(p,q,r)/19200$ using Table~\ref{tab:H}. Define $\operatorname{supp}(x):=\{j\in[5]:x_j=1\}$. For inputs $x,y$, let
\begin{equation}\label{eq:input-orbit-parameters}
 a=\min(|x|,|y|),\quad b=\max(|x|,|y|),\quad
 c=|\operatorname{supp}(x)\cap\operatorname{supp}(y)|,
\end{equation}
and set $C_{xy}=c_{abc}/307200$ using Table~\ref{tab:C}; all unlisted orbit coefficients are zero. In Table~\ref{tab:H}, a dash indicates an orbit type that does not occur among the indices of $H_1$. All three matrices $H_1,H_2,C$ are real symmetric and invariant under simultaneous permutations of the five coordinates. Every listed orbit type in Table~\ref{tab:C} contains one weight in $\{3,4\}$ and one weight outside it, so $C$ is supported on cross-label pairs, meaning pairs with $f(x)\ne f(y)$.

\begin{table}[H]\centering
\caption{Orbit coefficients for $H_1$ and $H_2$.}\label{tab:H}
\begin{tabular}{rrr}\toprule
$(p,q,r)$&$h_1$&$h_2$\\\midrule
$(0,0,0)$&-39160&-58360\\
$(0,1,0)$&1526355&1526355\\
$(0,2,0)$&-&-1298831\\
$(1,1,0)$&-1313423&-1313423\\
$(1,1,1)$&-23040&-42240\\
$(1,2,0)$&-&1021083\\
$(1,2,1)$&-&1526355\\
$(2,2,0)$&-&-2518478\\
$(2,2,1)$&-&-1320335\\
$(2,2,2)$&-&-33792\\\bottomrule
\end{tabular}
\end{table}

The nonzero coefficients of $C$ are listed in Table~\ref{tab:C}.
\begin{table}[H]\centering
\caption{Nonzero orbit coefficients for $C$.}\label{tab:C}
\begin{tabular}{rr}\toprule
$(a,b,c)$&$c_{abc}$\\\midrule
$(0,3,0)$&21447408\\
$(0,4,0)$&-31637000\\
$(1,3,0)$&-11962875\\
$(1,3,1)$&-16148508\\
$(1,4,0)$&16286060\\
$(1,4,1)$&24857047\\
$(2,3,0)$&5451300\\
$(2,3,1)$&6655097\\
$(2,3,2)$&7901158\\
$(2,4,1)$&-12204683\\
$(2,4,2)$&-15383580\\
$(3,5,3)$&10457368\\\bottomrule
\end{tabular}
\end{table}
All entries are specified by orbit data; no numerical sampling is involved.

The two orbit tables completely specify the certificate. We next explain how permutation symmetry reduces the verification and how the remaining Fourier entries are computed exactly.

The orbit tables above provide the complete rational data needed to reproduce the certificate verification. The proof itself reconstructs the certificate over $\mathbb Q$ and verifies strict positivity by exact rational decompositions and Sylvester's criterion. In particular, a small negative floating-point eigenvalue is never treated as zero, and numerical failure of one construction is never used as an impossibility argument. Symmetry reduction for semidefinite and SOS problems is developed by Gatermann and Parrilo \cite{GP}. Methods for recovering rational SOS decompositions from numerical output are studied by Peyrl and Parrilo \cite{PP}. Here the displayed rational certificate is verified directly
by exact arithmetic.

\subsection{\texorpdfstring{$S_5$}{S5}-orbit compression}
Let $S_5$ act on the input coordinates. For $\pi\in S_5$, let $U_\pi$ be the input permutation chosen so that $U_\pi\chi_S=\chi_{\pi(S)}$, and let $P_{t,\pi}$ act on subset indices by $P_{t,\pi}e_S=e_{\pi(S)}$, where $e_S$ is a standard basis vector. Extend $\pi$ by setting $\pi(0)=0$. Then
\begin{equation}\label{eq:permutation-action}
 U_\pi F_t=F_tP_{t,\pi},\qquad
 P_{t+1,\pi}T_{t,i}P_{t,\pi}^{\mathsf T}=T_{t,\pi(i)}.
\end{equation}
Since $H_t$ and $C$ are invariant, the slack matrices for nonzero indices are related by permutation congruence. It is therefore enough to check $i=0$ and $i=1$; the remaining indices follow by swapping coordinate $1$ with coordinate $i$.

The parameters $(p,q,r)$ classify subset-pair types up to
coordinate permutations and exchange of the two subsets.
Similarly, $(a,b,c)$ classifies input-pair types up to
coordinate permutations and exchange of the two inputs.
For fixed $a\le b$, the number of ordered pairs satisfying
$|x|=a$, $|y|=b$, and
$|\operatorname{supp}(x)\cap\operatorname{supp}(y)|=c$ is
\begin{equation}\label{eq:orbit-count}
	N_{abc}=\binom5a\binom ac\binom{5-a}{b-c}.
\end{equation}
When $a<b$, the pairs with $|x|=b$ and $|y|=a$ form a second
ordered orbit of the same size.
Thus a sorted type $(a,b,c)$ contains $2N_{abc}$ ordered pairs
when $a<b$, and $N_{aac}$ when $a=b$.
This explains the factor $2$ used for the nonzero types in
Table~\ref{tab:C}.

\subsection{Fourier matrix entries}
We now compute the Fourier transform of $C$ exactly, which determines the final-layer slack matrices. For $S,R\subseteq[5]$, write the ordered parameters $p=|S|$, $q=|R|$, and $r=|S\cap R|$; these $p,q$ are unrelated to the query-index notation $q_i$. Define
\begin{equation}\label{eq:W-polynomial}
\begin{aligned}
 W_{pqr}(u,v,w)={}&(1+u+v+uvw)^{5-p-q+r}(1-u+v-uvw)^{p-r}\\
 &\cdot(1-u-v+uvw)^{r}(1+u-v-uvw)^{q-r}.
\end{aligned}
\end{equation}
Let $\mathcal T$ denote the twelve orbit triples listed in Table~\ref{tab:C}. If $[u^av^bw^c]$ denotes coefficient extraction, then
\begin{equation}\label{eq:character-sum}
 \sum_{x,y}C_{xy}\chi_S(x)\chi_R(y)
 =\frac1{307200}\sum_{(a,b,c)\in\mathcal T}c_{abc}
 \left([u^av^bw^c]W_{pqr}+[u^bv^aw^c]W_{pqr}\right).
\end{equation}
Indeed, for one coordinate the four cases contribute
\begin{equation}\label{eq:coordinate-contributions}
1,\quad (-1)^{\one_{j\in S}}u,\quad (-1)^{\one_{j\in R}}v,\quad (-1)^{\one_{j\in S}+\one_{j\in R}}uvw,
\end{equation}
respectively. Grouping coordinates by membership in $S$ and $R$ gives $W_{pqr}$. Thus the formula is an exact finite polynomial identity, with no normalization ambiguity.

Let $\widehat Q=F_3^{\mathsf T}CF_3$, where the rows and columns are indexed by subsets in $\mathcal S_3$. Although $F=F_2$ in the terminal Gram formula, multiplication by a query character can raise the Fourier degree from two to three. Thus $\widehat Q$ contains every coefficient needed below. The ordered values of $p,q$ in $W_{pqr}$ are not sorted; they retain the order of $S,R$. Moreover,
\begin{equation}\label{eq:Z-elements}
 (Z_{2,i})_{S,R}=(H_2)_{S,R}+\widehat Q_{S\triangle q_i,R\triangle q_i}.
\end{equation}
This identity determines every final-layer entry once $\widehat Q$ is known. For $N_{abc}$ from \eqref{eq:orbit-count},
\begin{equation}\label{eq:Q-entries}
\begin{aligned}
 \widehat Q_{\varnothing,\varnothing}
 &=\frac{2}{307200}\sum_{(a,b,c)\in\mathcal T}N_{abc}c_{abc}=\frac{3043}{480},\\
 \widehat Q_{\{1\},\{1\}}
 &=\frac{2}{307200}\sum_{(a,b,c)\in\mathcal T}N_{abc}c_{abc}
 \left(1-\frac{2(a+b-2c)}5\right)=\frac{24}{5}.
\end{aligned}
\end{equation}
The factor in the second line is the average character product on the first coordinate within an orbit. Consequently $Z_{2,0}(\varnothing,\varnothing)=33/10$ and $5Z_{2,1}(\varnothing,\varnothing)=845/96$. Every other entry follows from the same coefficient-extraction identity, without enumerating floating-point eigenvalues.

\subsection{Positivity and permutation reduction}
The preceding formulas determine every entry of the final-layer slack matrices. We can therefore reduce their positivity to a small number of invariant blocks.

The matrices $H_t$ and $C$ are invariant under simultaneous permutations of the five input positions. If $P_{t,\pi}$ denotes the induced permutation on subset indices, then
\begin{equation}\label{eq:permutation-query-action}
 P_{t+1,\pi}T_{t,i}P_{t,\pi}^{\mathsf T}=T_{t,\pi(i)}.
\end{equation}
Consequently all nonzero-index slack matrices are permutation-congruent. The eighteen matrices reduce to the scalar representatives $Z_{0,0}=[499/480]$, $Z_{0,i}=[1/5]$, the identity matrix $Z_{1,0}=I_6$, and three nontrivial blocks
\begin{equation}\label{eq:slack-blocks}
 A=5Z_{1,1},\qquad B=Z_{2,0},\qquad D=5Z_{2,1}.
\end{equation}
The pair-space decomposition used to derive these blocks is given in Appendix~\ref{app:decomp}. Appendix~\ref{app:blocks} gives their complete quadratic-form decompositions. The five small matrices are $K_A$, $K_{B,0}$, $K_{B,1}$, $K_{D,0}$, and $K_{D,1}$; all have strictly positive leading principal minors, so Sylvester's criterion proves $A,B,D\succ0$. The remaining terms are positive multiples of squared norms. Hence all eighteen slack matrices are in fact positive definite, which proves the hypotheses of Lemma~\ref{lem:dual}. The version of Sylvester's criterion used here is stated in Appendix~\ref{app:sylvester}.

\begin{proposition}\label{prop:certificate}
The rational orbit tables above define a valid dual certificate: every $Z_{t,i}$ is real symmetric positive definite, and therefore $\QE(f)\ge4$.
\end{proposition}
\begin{proof}
The block decompositions and exact principal minors are given in Appendix~\ref{app:blocks}; permutation congruence transfers positivity from the six representatives to all indices. The support condition for $C$ was checked directly from the table. Lemma~\ref{lem:dual} then excludes all exact algorithms making at most three queries.
\end{proof}

\begin{proof}[Proof of Theorem~\ref{thm:main}]
By Proposition~\ref{prop:certificate}, \(\QE(f)\ge4\). The upper bound is inherited from the previously known result of Ambainis, Iraids, and Nagaj: Theorem~1 gives
\(\QE(\Exact^7_{3,4})=4\); in particular, it provides a four-query exact algorithm, which we use as a black box~\cite{AIN}. We now spell out the standard padding reduction. Set $k=3$ and pad the input by two fixed zero bits:
\begin{equation}\label{eq:padding-identity}
 f(x_1,\ldots,x_5)=\Exact^7_{3,4}(x_1,\ldots,x_5,0,0).
\end{equation}
For completeness, the standard padding simulation is as follows. A seven-bit phase query can be simulated coherently with at most one five-bit query by preserving the original index and routing it to $r(i)=i$ for $1\le i\le5$ and to $r(i)=0$ for $i=0,6,7$. Apply the five-bit query and reverse the routing. This works on arbitrary superpositions of indices and uses no extra query. Therefore $\QE(f)\le4$, completing the proof.
\end{proof}

\section{SOS Representation and Approximate-Error Lower Bound}\label{sec:consequences}
This section reviews a degree-three SOS representation of $f$ and derives an error lower bound from the rational certificate. The SOS identity is a pointwise algebraic fact, whereas the error bound follows from constraints on the query process. The definitions of SOS degree and the distinction between pointwise SOS identities and query-process constraints are summarized in Appendix~\ref{app:sos}.

\subsection{The symmetric polynomial and its SOS representation}
We first record the symmetric polynomial representation and then derive its fully symmetric SOS form. The elementary symmetric-polynomial calculation used here is recalled in Appendix~\ref{app:boolean}.

Let $e_j(x)=\sum_{|S|=j}\prod_{i\in S}x_i$. On the layer $|x|=w$, $e_j(x)=\binom{w}{j}$, so the six layer values determine the unique symmetric multilinear representation. The representation is
\begin{equation}\label{eq:poly}
 f(x)=e_3(x)-3e_4(x)+5e_5(x),
\end{equation}
because on weight $w$ the right side is $\binom w3-3\binom w4+5\binom w5$, equal to $f$ on all six layers. Thus $\deg f=5$.

For a nonnegative function \(g\) on the Boolean cube, let \(\sosdeg(g)\) be the least \(d\) for which \(g=\sum_\ell p_\ell^2\) on \(\{0,1\}^5\), with real multilinear polynomials \(p_\ell\) of degree at most \(d\). A degree-three SOS representation for this five-bit function is given by Xu and Qiu \cite[Eq.~(219)]{XQ}. We record a canonical fully \(S_5\)-symmetric form; the formal definition and the pointwise interpretation are given in Appendix~\ref{app:sos}. Put
\begin{equation}\label{eq:y-definition}
 y_{jk}(x)=\prod_{\ell\in[5]\setminus\{j,k\}}x_\ell,\quad Y=\sum_{j<k}y_{jk},\quad s_j=\sum_{k\ne j}y_{jk},
\end{equation}
where $y_{jk}=y_{kj}$ denotes the same unordered pair. Define
\begin{equation}\label{eq:h-definition}
 b_j=\frac{s_j-2Y/5}{3},\qquad h_{jk}=y_{jk}-\frac{Y}{10}-b_j-b_k.
\end{equation}
\begin{proposition}\label{prop:sos}
The function has SOS degree three, and the following is its fully $S_5$-symmetric representation:
\begin{equation}\label{eq:sos}
 f(x)=\frac54\sum_{j=1}^5 b_j(x)^2+\frac53\sum_{j<k}h_{jk}(x)^2.
\end{equation}
\end{proposition}
\begin{proof}
	Start with a degree-three SOS representation of $f$.
	Since $f(x)=0$ whenever $|x|\le2$, each polynomial in the SOS
	vanishes on these inputs. Evaluating successively at inputs of
	weights $0$, $1$, and $2$ shows that its constant, linear, and
	quadratic coefficients are zero. Hence every summand polynomial
	is a linear combination of the ten cubic monomials $y_{jk}$.
	
	Let $M_0\succeq0$ be the resulting Gram matrix in this basis,
	and let $P_\pi$ permute the pair coordinates according to
	$\pi\in S_5$. Define
	\[
	M:=\frac{1}{|S_5|}\sum_{\pi\in S_5}
	P_\pi^{\mathsf T}M_0P_\pi.
	\]
	Then $M\succeq0$ and $f(x)=y(x)^{\mathsf T}My(x)$ on the
	Boolean cube. Its entries have three values $a,b,c$ according
	as the indexing pairs are equal, intersect, or are disjoint.
	Evaluating the represented function on the weight-$3,4,5$
	layers gives
	\begin{equation}\label{eq:sos-coefficients}
		a=1,\qquad
		4a+12b=1,\qquad
		10a+60b+30c=0,
	\end{equation}
	so $(a,b,c)=(1,-1/4,1/6)$.
	Its eigenvalues on the constant, vertex, and row-sum-zero
	subspaces are $0,5/12,5/3$, respectively.
	
	The pair-coordinate decomposition gives
	\begin{equation}\label{eq:sos-projection}
		y_{jk}=\frac{Y}{10}+b_j+b_k+h_{jk},\qquad
		\sum_{j=1}^5 b_j=0,\qquad
		\sum_{j<k}(b_j+b_k)^2=3\sum_{j=1}^5 b_j^2.
	\end{equation}
	Moreover, the definition of $h_{jk}$ gives
	$\sum_{k\ne j}h_{jk}=0$ for every $j$.
	Thus the three displayed components are orthogonal:
	they are respectively the constant, vertex, and row-sum-zero
	components. The constant component is annihilated by $M$,
	while the two other components contribute
	\begin{equation}\label{eq:sos-eigenvalue-to-form}
		\begin{aligned}
			y^{\mathsf T}My
			&=\frac5{12}\sum_{j<k}(b_j+b_k)^2
			+\frac53\sum_{j<k}h_{jk}^2\\
			&=\frac54\sum_{j=1}^5b_j^2
			+\frac53\sum_{j<k}h_{jk}^2.
		\end{aligned}
	\end{equation}
	This proves \eqref{eq:sos}.
	Degree $2$ is impossible because a sum of squares of
	degree-two polynomials has multilinear degree at most four,
	whereas $\deg f=5$. Hence $\sosdeg(f)=3$.
	The representation above is a canonical fully
	$S_5$-symmetric form of the degree-three SOS identity.
\end{proof}

\subsubsection{Spectral form and a pointwise check}
We now verify the SOS identity directly on the six Hamming-weight layers and identify the spectrum of its Gram matrix.

The matrix $M$ defined above has diagonal entries $1$, entries $-1/4$ for intersecting pairs, and entries $1/6$ for disjoint pairs. Its eigenvalues on the constant, vertex, and row-sum-zero subspaces are, respectively,
\begin{equation}\label{eq:sos-checks}
 1+6(-1/4)+3(1/6)=0,\quad
 1-1/4-2(1/6)=5/12,\quad
 1-2(-1/4)+1/6=5/3.
\end{equation}
The identity \eqref{eq:sos} is $y^{\mathsf T}My$. If $|x|\le2$, all $y_{jk}$ vanish. If $|x|=3$, exactly one is one. If $|x|=4$, exactly four coordinates $y_{jk}$ equal $1$, and their indexing pairs all contain the unique zero
coordinate of $x$, giving $4+2\binom42(-1/4)=1$. If $|x|=5$, the all-ones vector lies in the zero-eigenvalue subspace, giving zero. Thus the identity holds on every Boolean input.

Equivalently, if $P_0,P_1,P_2$ are the projections onto the constant, vertex-induced zero-sum, and row-sum-zero subspaces, then
\begin{equation}\label{eq:sos-matrix}
 M=0\cdot P_0+\frac5{12}P_1+\frac53P_2.
\end{equation}
For example, at $x=(1,1,1,0,0)$ one has $y_{45}=1$ and all other $y_{jk}=0$; then $b_1=b_2=b_3=-2/15$, $b_4=b_5=1/5$, so $\sum_jb_j^2=2/15$, $\sum_{j<k}h_{jk}^2=1/2$, and the right side is $1/6+5/6=1$.

Lee, Prakash, de Wolf, and Yuen \cite{LeePrakashdeWolfYuen} study SOS degree for symmetric quadratic functions. The identity above concerns the present degree-five function and is verified independently. As a pointwise identity, it does not specify the inner products between state components at different inputs, nor the inter-layer recurrence in Lemma~\ref{lem:necessary}. Therefore $\sosdeg(f)=3$ does not imply $\QE(f)=3$; here $3=\sosdeg(f)<\QE(f)=4$.

\subsection{A quantitative consequence for approximate algorithms}
The SOS calculation is pointwise and does not establish feasibility of a three-query quantum process. We now return to the dual certificate of Section~\ref{sec:certificate} and derive a quantitative lower bound for approximate algorithms.

\begin{corollary}\label{cor:error}
For an algorithm $A$, let $Q(A)$ denote its worst-case number of oracle queries, and let $\varepsilon(A)=\max_{x\in X}\Pr[A(x)\ne f(x)]$. Define the optimal worst-case error with at most three queries by
\begin{equation}\label{eq:optimal-three-error}
 \varepsilon_3^\star(f):=\inf_{A:\,Q(A)\le3}\varepsilon(A).
\end{equation}
Every algorithm making at most three queries satisfies
\begin{equation}\label{eq:error}
 \varepsilon(A)\ge\frac{25600}{43779488192449}>5.84\times10^{-10},
\end{equation}
and consequently
$\varepsilon_3^\star(f)\ge25600/43779488192449$.
\end{corollary}
\begin{proof}
The Gram recurrence holds for an approximate algorithm as well. Pairing it with the same certificate and using $Z_{t,i},A_{t,i}\succeq0$ gives $\sum_{t,i}\Tr(Z_{t,i}A_{t,i})\ge0$, hence $-1+\Tr(CG)\ge0$, that is, $\Tr(CG)\ge1$. Let $E_0,E_1$ be the final POVM with $E_0+E_1=I$. For $f(x)\ne f(y)$, Cauchy--Schwarz gives
\begin{equation}\label{eq:cross-label-bound}
 |G_{xy}|\le\sum_{b=0}^1\|E_b^{1/2}\psi_x\|\,\|E_b^{1/2}\psi_y\|\le2\sqrt{\varepsilon(A)},
\end{equation}
because for each output value $b$, one of the two factors is an error amplitude at most $\sqrt{\varepsilon(A)}$, while the other is at most $1$. Since $C$ is supported only on cross-label pairs, and every nonzero orbit in Table~\ref{tab:C} has $a<b$, each unordered cross-label input pair contributes the two ordered entries $C_{xy}$ and $C_{yx}$. Define its entrywise absolute sum
\begin{equation}\label{eq:certificate-norm-orbit}
 L:=\sum_{x,y}|C_{xy}|
 =\frac{2}{307200}\sum_{(a,b,c)\in\mathcal T}N_{abc}|c_{abc}|
 =\frac{6616607}{320}.
\end{equation}
Then $|\Tr(CG)|\le L\max_{f(x)\ne f(y)}|G_{xy}|\le2L\sqrt{\varepsilon(A)}$, so
\begin{equation}\label{eq:certificate-norm-bound}
 1\le|\Tr(CG)|\le2L\sqrt{\varepsilon(A)}.
\end{equation}
Thus $\varepsilon(A)\ge1/(4L^2)=25600/43779488192449$. Taking the infimum over all algorithms with at most three queries gives
$\varepsilon_3^\star(f)\ge25600/43779488192449>0$. The trace is real because $C$ and $G$ are Hermitian; no complex-phase issue is hidden in the inequality.

The displayed constant is therefore a valid lower bound on the true optimal three-query error: it may be strictly smaller than $\varepsilon_3^\star(f)$, but it cannot be larger, and we do not claim equality or optimality. The bound depends on the present certificate norm $L$, but it is a robust consequence of the same exact certificate: initialization and the Gram recurrence remain valid for approximate algorithms, the support of $C$ still connects only cross-label pairs, and each cross-label inner product is bounded by two measurement branches, one of which carries an error factor. No symmetry assumption is made about the approximate algorithm.
\end{proof}

\section{Conclusion}\label{sec:conclusion}
We have proved \(\QE(\Exact^5_{3,4})=4\). The main new ingredient is a global rational dual certificate for the necessary Fourier--Gram constraints of every three-query process. It imposes no restriction on workspace dimension, amplitudes, intermediate measurements, or permutation symmetry, and it is finite and rational, so the lower bound is checkable by exact arithmetic rather than by numerical optimization. The upper bound is inherited from the previously known four-query algorithm for \(\Exact^7_{3,4}\) via the standard padding reduction. The degree-three SOS representation is a separate pointwise algebraic fact and does not imply a three-query algorithm; it gives a concrete separation between the SOS degree of $f$ and its exact quantum query complexity, $\sosdeg(f)=3<\QE(f)=4$. The same certificate also yields the valid, though not necessarily tight, bound
\[
\varepsilon_3^\star(f)\ge
\frac{25600}{43779488192449}
\]
for the optimal worst-case error of algorithms using at most three queries. Future work may seek the optimal value of
$\varepsilon_3^\star(f)$ and investigate analogous rational certificates for other adjacent-weight instances.

\begin{appendices}
\section{Background and notation}\label{app:background}
This section collects standard definitions and elementary facts used in the proof. It is intended as a short guide for readers who do not routinely work with quantum query algorithms, Fourier characters, Gram matrices, or semidefinite certificates. None of the background facts in this section is claimed as a new result. The later appendices contain paper-specific derivations, certificate calculations, and exact verification details.

\subsection{Boolean functions and multilinear polynomials}\label{app:boolean}
The Boolean cube is $\{0,1\}^n$, and the Hamming weight of $x$ is $|x|=\sum_{j=1}^n x_j$. A polynomial is multilinear when every variable has exponent at most one. Every real-valued function on $\{0,1\}^n$ has a unique multilinear representation: reduce powers using $x_j^2=x_j$, and then use the $2^n$ evaluations on the cube to determine the $2^n$ monomial coefficients \cite{BBCMW,NisanSzegedy}. For a symmetric function, the value depends only on $|x|$. The elementary symmetric polynomial
\begin{equation}\label{eq:bg-elementary-symmetric}
 e_j(x)=\sum_{\substack{S\subseteq[n]\\ |S|=j}}\prod_{i\in S}x_i
\end{equation}
satisfies $e_j(x)=\binom{|x|}{j}$ on the layer of weight $|x|$. This is why the six layer values determine the representation in \eqref{eq:poly}.

\subsection{The query model and purification}\label{app:query-model}
A $T$-query quantum algorithm alternates input-independent unitaries with calls to an oracle. After padding an algorithm that uses fewer than $T$ calls with identity calls, its final state can be written as
\begin{equation}\label{eq:bg-query-computation}
 |\psi_T(x)\rangle=U_T O_x U_{T-1}O_x\cdots O_x U_1O_xU_0|\psi_{\mathrm{in}}\rangle .
\end{equation}
The workspace may be arbitrarily large, and the unitaries may depend on the algorithm but not on $x$. The usual bit oracle and the phase oracle are equivalent in the standard query model \cite{BSS,GrilloMarquezino}: in the answer-qubit Hadamard basis the bit oracle leaves $|+\rangle$ unchanged and multiplies $|-\rangle$ by $(-1)^{x_i}$, and routing the two components to a virtual index $0$ and the queried index respectively yields \eqref{eq:phase-query}. The original index and the answer-branch label are retained in auxiliary registers, so the routing is reversible. The auxiliary registers are uncomputed after the query. Thus the conversion uses the same number of oracle calls in both directions.

Randomness and intermediate measurements do not enlarge the model. Attach a register containing the random seed and measurement outcomes, keep it coherent, and defer the final measurement; this purification produces a unitary computation whose final measurement has the same output distribution. An early-stopping branch is completed with the virtual index-$0$ identity query. Thus the Gram constraints apply to every finite-workspace query algorithm, including algorithms with internal randomness or intermediate measurements.

\subsection{Fourier characters}\label{app:fourier}
For $S\subseteq[n]$, the Boolean Fourier character is
\begin{equation}\label{eq:bg-character}
 \chi_S(x)=(-1)^{\sum_{j\in S}x_j}.
\end{equation}
Characters multiply by symmetric difference and are orthogonal under the uniform sum:
\begin{equation}\label{eq:bg-fourier-orthogonality}
 \chi_S(x)\chi_R(x)=\chi_{S\triangle R}(x),\qquad
 \sum_{x\in\{0,1\}^n}\chi_S(x)\chi_R(x)=
 \begin{cases}
 2^n,&S=R,\\
 0,&S\ne R.
 \end{cases}
\end{equation}
The use of these characters as a query-adapted basis is standard in quantum query complexity \cite{BBCMW,BSS}. For the second identity, if $S\ne R$, choose $j\in S\triangle R$ and pair every input $x$ with the input obtained by flipping $x_j$. The two paired summands have opposite signs. If $S=R$, every summand is $1$. Consequently the character-evaluation matrix, divided by $2^{n/2}$, has orthonormal columns. A phase query multiplies a character by $\chi_{\{i\}}$, which sends the Fourier label $S$ to $S\triangle\{i\}$ and raises its degree by at most one.

\subsection{Gram matrices and positive semidefiniteness}\label{app:gram}
Given vectors $v_1,\ldots,v_m$ in a complex inner-product space, their Gram matrix is
\begin{equation}\label{eq:bg-gram}
 A_{jk}=\langle v_j,v_k\rangle .
\end{equation}
Gram matrices and their use in query-complexity semidefinite formulations are standard; see, for example, Barnum, Saks, and Szegedy \cite{BSS}. For every $z\in\mathbb C^m$, $z^*Az$ is a squared norm, so $A$ is Hermitian positive semidefinite (PSD). In the query proof, the vectors are the Fourier coefficient vectors $v_{t,i,S}$; hence their Gram matrices retain all pairwise input information even when the workspace dimension is unrestricted.

If $A$ and $Z$ are PSD matrices of the same size, then
\begin{equation}\label{eq:bg-trace-positivity}
 \operatorname{Tr}(ZA)=\operatorname{Tr}\!\left(Z^{1/2}AZ^{1/2}\right)\ge0.
\end{equation}
This elementary trace inequality is the only positivity fact needed when the algorithm's Gram matrices are complex Hermitian and the certificate matrices are real symmetric.

\subsection{Measurements and exact output}\label{app:povm}
A two-outcome measurement is described by effects $E_0,E_1\succeq0$ with $E_0+E_1=I$; this is the standard output-measurement formalism used in the query model \cite{BSS}. Exact computation means
\begin{equation}\label{eq:bg-exact-povm}
 \langle\psi_x|E_{f(x)}|\psi_x\rangle=1,\qquad
 \langle\psi_x|E_{1-f(x)}|\psi_x\rangle=0
 \quad\text{for every }x.
\end{equation}
For a PSD effect, expectation zero implies $E_b^{1/2}|\psi_x\rangle=0$. Therefore, if $f(x)\ne f(y)$, insert $I=E_0+E_1$ between the two final states and use Cauchy--Schwarz on each branch:
\begin{equation}\label{eq:bg-orthogonal-final}
 \langle\psi_x|\psi_y\rangle
 =\sum_{b=0}^1
 \langle E_b^{1/2}\psi_x,E_b^{1/2}\psi_y\rangle=0.
\end{equation}
This is the output-orthogonality condition used in Lemma~\ref{lem:necessary}. For bounded error, the same calculation gives the cross-label estimate in \eqref{eq:cross-label-bound}.

\subsection{Dual certificates as separating functionals}\label{app:dual}
The recurrence in Lemma~\ref{lem:necessary} defines a finite-dimensional necessary feasibility region for three-query algorithms, in the spirit of the Gram-matrix and semidefinite formulations of quantum query complexity \cite{BSS}. A certificate chooses fixed matrices $H_t$ and $C$ so that the associated slacks $Z_{t,i}$ are PSD and $C$ vanishes on equal-label input pairs. For every feasible Gram family, the recurrence makes the following expression telescope:
\begin{equation}\label{eq:bg-dual-telescope}
 \sum_{t=0}^{2}\sum_i\operatorname{Tr}(Z_{t,i}A_{t,i})
 =-1+\operatorname{Tr}(CG).
\end{equation}
The left side is nonnegative by \eqref{eq:bg-trace-positivity}. Exact output makes $G_{xy}=0$ on all pairs on which $C$ is supported, so the right side is $-1$. This contradiction separates the necessary three-query region from the exact-output condition. The argument checks one explicit rational functional and does not require strong duality, a converse characterization of algorithms, or a restriction to symmetric algorithms.

\subsection{Sylvester's criterion}\label{app:sylvester}
For a real symmetric $r\times r$ matrix $K$, Sylvester's criterion states
\begin{equation}\label{eq:bg-sylvester}
 K\succ0\quad\Longleftrightarrow\quad
 \det K_{[1:j],[1:j]}>0\quad\text{for every }j=1,\ldots,r,
\end{equation}
where $K_{[1:j],[1:j]}$ is the leading $j\times j$ principal submatrix. The proof uses an $LDL^{\mathsf T}$ elimination without pivoting: the successive pivots are ratios of consecutive leading principal minors, so positivity of all of them is equivalent to positive definiteness. In this paper the criterion is applied only to the five explicitly displayed rational matrices in Appendix~\ref{app:blocks}; all remaining terms in the quadratic-form decompositions are visibly nonnegative squares.

\subsection{SOS degree and its limitation}\label{app:sos}
For a nonnegative function $g$ on the Boolean cube, its SOS degree is the least $d$ for which there are real multilinear polynomials $p_\ell$ of degree at most $d$ satisfying the pointwise identity
\begin{equation}\label{eq:bg-sos-degree}
 g(x)=\sum_\ell p_\ell(x)^2\qquad\text{for every }x\in\{0,1\}^n.
\end{equation}
This is an algebraic identity evaluated separately at each input. It constrains the values of $g$ but does not specify inner products between final states for different inputs, nor does it impose the layer-by-layer recurrence of the query Gram matrices. A degree-three SOS representation can therefore coexist with a four-query lower bound. The degree-three SOS representation underlying Section~\ref{sec:consequences} is given in \cite[Eq.~(219)]{XQ}. The fully $S_5$-symmetric form displayed here is obtained by averaging its Gram matrix over coordinate permutations. For further SOS-degree results in the symmetric setting, see \cite{LeePrakashdeWolfYuen}. 

\section{Fourier--Gram and dual proofs}
\label{app:proof-details}
This appendix records the expanded derivations specific to the present certificate that are omitted from the main text. They are included for auditability; the main theorem uses the concise statements and proofs in Sections~\ref{sec:gram}--\ref{sec:certificate}.

\subsection{Proof of the Gram recurrence}
For reference, the Fourier expansion used in the seven-step proof is
\begin{equation}\label{eq:state-fourier-expansion}
 |a_{t,i}(x)\rangle=\sum_{S\in\cS_t}\chi_S(x)|v_{t,i,S}\rangle.
\end{equation}

For completeness, the same derivation is organized below into seven labeled steps; this is an expanded restatement for auditability.

\textbf{Step 1: uniform query form.} Purify the initial state, all input-independent randomness, and every intermediate measurement record. An early-stopping branch is completed with the identity query at virtual index $0$, so every branch has three query slots and the only input-dependent operation is $O_x$.

\textbf{Step 2: low-degree Fourier expansion.} Write the state before query $t+1$ as
\begin{equation}\label{eq:state-before-query}
 |\phi_t(x)\rangle=\sum_{i=0}^{5}|i\rangle|a_{t,i}(x)\rangle.
\end{equation}
The expansion \eqref{eq:state-fourier-expansion} follows by induction. At $t=0$ the state is independent of $x$. If it holds at level $t$, the phase query multiplies the $i$th branch by $\chi_{q_i}(x)$ and sends each character $\chi_S$ to $\chi_{S\triangle q_i}$; the following input-independent unitary only forms linear combinations. This proves the degree bound for every algorithm, not just a searched ansatz.

\textbf{Step 3: coefficient Gram matrices.} Define
\begin{equation}\label{eq:gram-entry-expanded}
 (A_{t,i})_{S,R}=\langle v_{t,i,S}\mid v_{t,i,R}\rangle.
\end{equation}
For every complex vector $z$,
\begin{equation}\label{eq:gram-psd}
 z^*A_{t,i}z=\left\|\sum_Sz_Sv_{t,i,S}\right\|^2\ge0.
\end{equation}
Thus $A_{t,i}$ is a complex Hermitian PSD matrix, even when the algorithm uses arbitrary complex amplitudes. Its dimension is $d_t$, independent of the workspace dimension.

\textbf{Step 4: initial normalization.} At level $t=0$, only $S=\varnothing$ occurs. Hence
\begin{equation}\label{eq:initial-norm}
 \langle\phi_0(x)\mid\phi_0(x)\rangle
 =\sum_{i=0}^{5}\|v_{0,i,\varnothing}\|^2=1,
\end{equation}
which is precisely $\sum_iA_{0,i}=[1]$.

\textbf{Step 5: the layer recurrence.} After the query with index $i$, the coefficient of character $R$ is
\begin{equation}\label{eq:queried-coefficients}
 |\widetilde v_{t,i,R}\rangle
 =\sum_{S\in\cS_t}(T_{t,i})_{R,S}|v_{t,i,S}\rangle.
\end{equation}
Therefore the post-query coefficient Gram matrix is
$\widetilde A_{t,i}=T_{t,i}A_{t,i}T_{t,i}^{\mathsf T}$. The next input-independent unitary preserves the inner product of the complete state for every pair $(x,y)$ and merely redistributes these coefficients among the next query-index branches. Summing those branches yields \eqref{eq:recurrence}.

The same argument can be written as a kernel identity. Before and after the input-independent unitary, the value-space Gram matrices are respectively
\begin{equation}\label{eq:kernel-identity}
 F_{t+1}\left(\sum_iT_{t,i}A_{t,i}T_{t,i}^{\mathsf T}\right)F_{t+1}^{\mathsf T}
 \quad\text{and}\quad
 F_{t+1}\left(\sum_iA_{t+1,i}\right)F_{t+1}^{\mathsf T}.
\end{equation}
They are equal for every pair of inputs. Since $F_{t+1}^{\mathsf T}F_{t+1}=32I$, the matrix $F_{t+1}$ has full column rank; multiplying on the left by $F_{t+1}^{\mathsf T}/32$ and on the right by $F_{t+1}/32$ therefore recovers the coefficient recurrence. This is why the proof uses all pairwise kernels rather than only diagonal norms.

\textbf{Step 6: terminal Gram matrix.} Before the third query, the $i$th branch contributes the value-space Gram kernel $F A_{2,i}F^{\mathsf T}$. The query conjugates this kernel by $D_i$, and orthogonality of the query-index branches makes the final Gram matrix the sum in \eqref{eq:G}. Any later input-independent unitary leaves that Gram matrix unchanged.

\textbf{Step 7: exact output orthogonality.} Let the final POVM be $E_0,E_1$, with $E_0+E_1=I$. If $f(x)=0$ and $f(y)=1$, exactness gives
\begin{equation}\label{eq:measurement-probabilities}
 \langle\psi_x|E_0|\psi_x\rangle=1,\qquad
 \langle\psi_y|E_0|\psi_y\rangle=0.
\end{equation}
For a PSD operator $B$, $\langle v|B|v\rangle=0$ implies $Bv=0$. Apply this first to $B=I-E_0$ and then to $B=E_0$ to obtain $E_0\psi_x=\psi_x$ and $E_0\psi_y=0$. Hence
$G_{xy}=\langle\psi_x|E_0|\psi_y\rangle=0$. This completes the proof of the necessary condition.

\subsection{Proof of the dual certificate lemma}
It is useful to view $H_t$ as a potential on the state before the $(t+1)$st query. The difference
\begin{equation}\label{eq:terminal-slack-expanded}
 H_t-T_{t,i}^{\mathsf T}H_{t+1}T_{t,i}
\end{equation}
compares the potential before a query with the potential after its coefficient-space action. The terminal slack
$H_2+F^{\mathsf T}D_iCD_iF$ incorporates the final Gram constraint.

For every PSD pair $Z_{t,i},A_{t,i}$,
\begin{equation}\label{eq:dual-nonnegative}
 \operatorname{Tr}(Z_{t,i}A_{t,i})
 =\operatorname{Tr}(Z_{t,i}^{1/2}A_{t,i}Z_{t,i}^{1/2})\ge0,
\end{equation}
with no commutativity assumption. Summing the first two layers and using cyclicity gives
\begin{equation}\label{eq:layer-trace}
 \sum_i\operatorname{Tr}(Z_{t,i}A_{t,i})
 =\operatorname{Tr}\left(H_t\sum_iA_{t,i}\right)
 -\operatorname{Tr}\left(H_{t+1}\sum_iT_{t,i}A_{t,i}T_{t,i}^{\mathsf T}\right).
\end{equation}
At the terminal layer,
\begin{equation}\label{eq:terminal-trace}
 \sum_i\operatorname{Tr}(F^{\mathsf T}D_iCD_iFA_{2,i})
 =\operatorname{Tr}\left(C\sum_iD_iFA_{2,i}F^{\mathsf T}D_i\right)
 =\operatorname{Tr}(CG).
\end{equation}
Thus the complete telescoping identity is
\begin{equation}\label{eq:dual-value}
 \sum_{t=0}^{2}\sum_i\operatorname{Tr}(Z_{t,i}A_{t,i})
 =-1+\operatorname{Tr}(CG).
\end{equation}
Since $\operatorname{Tr}(CG)=\sum_{x,y}C_{xy}G_{yx}$, the support condition on $C$ and cross-label orthogonality of $G$ imply
$\operatorname{Tr}(CG)=0$. Therefore
\begin{equation}\label{eq:dual-contradiction}
 \sum_{t=0}^{2}\sum_i\operatorname{Tr}(Z_{t,i}A_{t,i})=-1.
\end{equation}
This contradicts \eqref{eq:dual-nonnegative}. The certificate excludes the whole necessary feasible region, rather than one circuit ansatz or one SOS decomposition.

This is a finite-dimensional dual infeasibility proof. It does not invoke strong SDP duality or assume real amplitudes. The certificate is real and rational, whereas an arbitrary algorithm may have complex Hermitian Gram matrices.

\section{Orthogonal decomposition on pairs}\label{app:decomp}
This section provides the invariant decomposition used in the block calculations of Appendix~\ref{app:blocks}.

\begin{lemma}\label{lem:pair}
Let $m\ge3$. We use the convention $v_{jk}=v_{kj}$ and $h_{jk}=h_{kj}$ for $j\ne k$, so that row sums may be written without ordering restrictions. Every vector $(v_{jk})_{1\le j<k\le m}$ has a unique decomposition
\begin{equation}\label{eq:pair}
 v_{jk}=\gamma+b_j+b_k+h_{jk},\qquad \sum_jb_j=0,\qquad \sum_{k\ne j}h_{jk}=0.
\end{equation}
The three components are mutually orthogonal, and
\begin{equation}\label{eq:bnorm}
 \sum_{j<k}(b_j+b_k)^2=(m-2)\sum_jb_j^2.
\end{equation}
Their dimensions are $1$, $m-1$, and $\binom m2-m$.
\end{lemma}
\begin{proof}
Let $s_j=\sum_{k\ne j}v_{jk}$ and $\bar s=m^{-1}\sum_js_j$. Set
\begin{equation}\label{eq:decomposition-parameters}
 \gamma=\frac{\bar s}{m-1},\qquad b_j=\frac{s_j-\bar s}{m-2},\qquad h_{jk}=v_{jk}-\gamma-b_j-b_k.
\end{equation}
Then $\sum_jb_j=0$ and each $h$ row sum is zero. Conversely, any decomposition with these constraints satisfies $s_j=(m-1)\gamma+(m-2)b_j$, proving uniqueness. Orthogonality follows by summing the zero-sum conditions, and \eqref{eq:bnorm} follows from $2\sum_{j<k}b_jb_k=-\sum_jb_j^2$. The map $b\mapsto(b_j+b_k)$ is injective on the zero-sum subspace. The row-sum map has rank $m$, so its kernel has dimension $\binom m2-m$.
\end{proof}

\begin{example}\label{ex:pair}
For $m=4$, take $v_{12}=1$ and all other pair coordinates zero. The decomposition gives
\begin{equation}\label{eq:decomposition-example}
 \gamma=1/6,\qquad b=(1/4,1/4,-1/4,-1/4),
\end{equation}
and, in the order $12,13,14,23,24,34$,
\begin{equation}\label{eq:harmonic-example}
 h=(1/3,-1/6,-1/6,-1/6,-1/6,1/3).
\end{equation}
For instance $v_{12}=1/6+1/4+1/4+1/3=1$ and $v_{13}=1/6+1/4-1/4-1/6=0$; every row sum of $h$ is zero.
\end{example}

For later use, let $u=(u_j)$ and $v=(v_j)$ be arbitrary point-coordinate vectors with $u_j=\beta+a_j$, $v_j=\delta+b_j$, and zero-sum $a,b$. Let $w_{jk}=\eta+c_j+c_k+h_{jk}$ be pair coordinates with zero-sum $c$ and row-sum-zero $h$. If a point matrix has diagonal/off-diagonal entries $p,q$, then
\begin{equation}\label{eq:point}
 p\sum_ju_jv_j+q\sum_{j\ne k}u_jv_k
 =m(p+(m-1)q)\beta\delta+(p-q)\sum_ja_jb_j.
\end{equation}
If a point--pair block has entry $R_{j,kl}=r$ when $j\in\{k,l\}$ and $R_{j,kl}=s$ otherwise, then
\begin{equation}\label{eq:point-pair}
 \sum_j\sum_{k<l}R_{j,kl}u_jw_{kl}
 =\binom m2(2r+(m-2)s)\beta\eta+(m-2)(r-s)\sum_ja_jc_j.
\end{equation}
If a pair matrix has entries $d,e,g$ for equal, intersecting, and disjoint pairs, then
\begin{align}
 w^{\mathsf T}Mw={}&\binom m2\left(d+2(m-2)e+\binom{m-2}{2}g\right)\eta^2\\
 &+(m-2)\bigl(d+(m-4)e-(m-3)g\bigr)\sum_jc_j^2
 +(d-2e+g)\sum_{j<k}h_{jk}^2.\label{eq:pairs-form}
\end{align}
These formulas are obtained by separating constant, vertex, and row-sum-zero components and by counting equal, intersecting, and disjoint pairs; they are the elementary block calculations used below. In particular,
\begin{equation}\label{eq:pair-sum-identity}
 \sum_j a_j\sum_{k\ne j}(c_j+c_k)=(m-2)\sum_ja_jc_j,
\end{equation}
while the contribution from pairs avoiding $j$ is the negative of this quantity. For a fixed pair, the sum over intersecting $h$-coordinates is $-2h_{jk}$ and the sum over disjoint coordinates is $h_{jk}$, which yields the coefficient $d-2e+g$ in \eqref{eq:pairs-form}. If the scalar-to-point and scalar-to-pair blocks have constant entries $\ell$ and $\kappa$, their contributions to the mean coordinates are $m\ell$ and $\binom{m}{2}\kappa$; in a symmetric quadratic form, an off-diagonal block is counted twice.

\paragraph{Invariant-subspace interpretation.}
The pair-coordinate decomposition can also be expressed
in terms of invariant subspaces.

The pair-coordinate space is the space of edge functions
on the complete graph $K_m$.
Let $V_{\mathrm{pt}}$ consist of all vectors
$(b_j+b_k)_{j<k}$ with $b\in\mathbb R^m$ and
$\sum_{j=1}^m b_j=0$.
Let $V_{\mathrm{harm}}$ consist of all pair vectors
$(h_{jk})_{j<k}$ satisfying
$\sum_{k\ne j}h_{jk}=0$ for every $j$,
with the convention $h_{jk}=h_{kj}$.
With $\mathbf 1$ denoting the all-ones vector,
\begin{equation}\label{eq:coordinate-decomposition}
 \mathbb R^{\binom m2}=\operatorname{span}\{\mathbf 1\}\oplus V_{\mathrm{pt}}\oplus V_{\mathrm{harm}}.
\end{equation}
These three subspaces have dimensions
$1$, $m-1$, and $\binom m2-m$, respectively.
Any matrix depending only on pair intersection size
preserves them. This is the reason the large slack
matrices reduce to small blocks and may also be viewed
as the elementary $S_m$-representation decomposition
on two-subsets.

\section{Positivity and exact verification}\label{app:blocks}
We now verify positivity of the three nontrivial slack blocks.

In this appendix, the uppercase block symbols $A,B,D$ denote slack blocks and should not be confused with the query-diagonal matrices $D_i$. We set $A=5Z_{1,1}$, $B=Z_{2,0}$, and $D=5Z_{2,1}$. Their entries are obtained from the orbit tables and \eqref{eq:character-sum}.

\subsection{\texorpdfstring{Block $A$}{Block A}}
We first analyze the $6\times6$ slack block $A$.

Order coordinates as $\varnothing,\{1\},\ldots,\{5\}$, write $z=(\alpha,\beta,u_2,u_3,u_4,u_5)$, and put $u_j=\gamma+r_j$ with $\sum_{j=2}^5r_j=0$. Then
\begin{equation}\label{eq:A-form}
 z^{\mathsf T}Az=\frac{77}{96}\alpha^2+(\beta,\gamma)K_A(\beta,\gamma)^{\mathsf T}+\sum_{j=2}^5r_j^2,
\end{equation}
where
\begin{equation}\label{eq:block-a-matrix}
 K_A=\begin{pmatrix}\frac{883}{96}&-\frac{76}{5}\\-\frac{76}{5}&\frac{164}{5}\end{pmatrix}.
\end{equation}
The empty-set diagonal of $A$ is $77/96$ and its cross entries with singleton coordinates are zero. The singleton diagonal entries are $883/96,14/5,14/5,14/5,14/5$; the cross entry between coordinate $\{1\}$ and any other singleton is $-19/5$, and between two of the remaining singletons is $9/5$. Hence the zero-sum coefficient is $14/5-9/5=1$, the mean coefficient is $4(14/5+3(9/5))=164/5$, and the cross term is $4(-19/5)=-76/5$.

\subsection{\texorpdfstring{Block $B$}{Block B}}
We next analyze the block $B$, whose coordinates consist of the empty set, singleton subsets, and pair subsets.

Order coordinates as the empty set, singletons, and unordered pairs. The independent entry types are listed in Table~\ref{tab:B}.
\begin{table}[H]\centering
\caption{Entry types of $B$.}\label{tab:B}
\begin{tabular}{rr@{\qquad\qquad}rr}\toprule
$(p,q,r)$&$B_{S,R}$&$(p,q,r)$&$B_{S,R}$\\\midrule
$(0,0,0)$&33/10&$(1,2,0)$&-23/120\\
$(0,1,0)$&-1/50&$(1,2,1)$&-37/400\\
$(0,2,0)$&-181/200&$(2,2,0)$&529/300\\
$(1,1,0)$&-1/5&$(2,2,1)$&-481/600\\
$(1,1,1)$&13/5&$(2,2,2)$&141/25\\\bottomrule
\end{tabular}
\end{table}
Put $\beta=\frac15\sum_{j=1}^5u_j$ and
$a_j=u_j-\beta$, so that $\sum_{j=1}^5a_j=0$.
Apply Lemma~\ref{lem:pair} to the pair coordinates to write
$v_{jk}=\gamma+b_j+b_k+h_{jk}$, where
$\sum_{j=1}^5b_j=0$ and
$\sum_{k\ne j}h_{jk}=0$ for every $j$.
Then
\begin{align}\label{eq:B-form}
 z^{\mathsf T}Bz={}&(\alpha,\beta,\gamma)K_{B,0}(\alpha,\beta,\gamma)^{\mathsf T}
 +\sum_{j=1}^5(a_j,b_j)K_{B,1}(a_j,b_j)^{\mathsf T}
 +\frac{1351}{150}\sum_{j<k}h_{jk}^2,
\end{align}
with
\begin{equation}\label{eq:block-b-matrices}
K_{B,0}=\begin{pmatrix}33/10&-1/10&-181/20\\-1/10&9&-38/5\\-181/20&-38/5&306/5\end{pmatrix},\qquad
K_{B,1}=\begin{pmatrix}14/5&119/400\\119/400&787/200\end{pmatrix}.
\end{equation}
For example, the mixed entry in $K_{B,1}$ is $3((-37/400)-(-23/120))=119/400$, its second diagonal entry is $3(141/25-481/600-2(529/300))=787/200$, and the row-sum-zero coefficient is $141/25-2(-481/600)+529/300=1351/150$. The dimensions $3+2\cdot4+5=16$ account for every coordinate.

\subsection{\texorpdfstring{Block $D$}{Block D}}
Finally, we analyze the block $D$, whose invariant decomposition involves the largest coefficient block.

Let $J=\{2,3,4,5\}$. Use coordinates
$\alpha,\beta,u_j,v_j,w_{jk}$ for
$\varnothing,\{1\},\{j\},\{1,j\},\{j,k\}$,
respectively, with $j,k\in J$ and $j<k$ for pair coordinates.
Put
$\gamma=\frac14\sum_{j\in J}u_j$,
$\delta=\frac14\sum_{j\in J}v_j$,
$a_j=u_j-\gamma$, and $b_j=v_j-\delta$.
Thus $\sum_{j\in J}a_j=\sum_{j\in J}b_j=0$.
Apply Lemma~\ref{lem:pair} on the index set $J$ to write
$w_{jk}=\eta+c_j+c_k+h_{jk}$, where
$\sum_{j\in J}c_j=0$ and
$\sum_{\substack{k\in J\\k\ne j}}h_{jk}=0$
for every $j\in J$.
The complete list of independent entry types of $D$ is given in Table~\ref{tab:D}. Different letters denote distinct elements of $J$; thus $(jk,jl)$ uses three distinct indices and represents intersecting pairs, whereas $(jk,lm)$ uses four distinct indices and represents disjoint pairs. In the pair entries, the first/second/third values correspond respectively to equal, intersecting, and disjoint pairs, or to the indicated membership cases.
\begin{table}[H]\centering\small
\caption{Entry types of $D$.}\label{tab:D}
\begin{tabular}{ll}\toprule
Entry type & Value\\\midrule
$(\alpha,\alpha),(\alpha,\beta),(\beta,\beta)$ & $845/96,-1/10,1987/96$\\
$(\alpha,u_j),(\alpha,v_j),(\alpha,w_{jk})$ & $-37/80,14/5,-5/8$\\
$(\beta,u_j),(\beta,v_j),(\beta,w_{jk})$ & $-333/40,-1/10,11/3$\\
$(u_j,u_j),(u_j,u_k)$ & $26,-53/24$\\
$(u_j,v_j),(u_j,v_k)$ & $-37/80,-23/24$\\
$(u_j,w_{jk}),(u_j,w_{kl})$ & $15569/120,-5317/40$\\
$(v_j,v_j),(v_j,v_k)$ & $76/5,-14/5$\\
$(v_j,w_{jk}),(v_j,w_{kl})$ & $-249/40,5671/960$\\
$(w_{jk},w_{jk}),(w_{jk},w_{jl}),(w_{jk},w_{lm})$ & $51129/20,59/80,-76619/30$\\\bottomrule
\end{tabular}
\end{table}

The quadratic form is
\begin{align}\label{eq:D-form}
 z^{\mathsf T}Dz={}&(\alpha,\beta,\gamma,\delta,\eta)K_{D,0}(\alpha,\beta,\gamma,\delta,\eta)^{\mathsf T}\\
 &+\sum_{j\in J}(a_j,b_j,c_j)K_{D,1}(a_j,b_j,c_j)^{\mathsf T}
 +\frac{121}{120}\sum_{\{j,k\}\subseteq J}h_{jk}^2,\nonumber
\end{align}
where
\begin{equation}\label{eq:block-d0-matrix}
K_{D,0}=\begin{pmatrix}
845/96&-1/10&-37/20&56/5&-15/4\\
-1/10&1987/96&-333/10&-2/5&22\\
-37/20&-333/10&155/2&-267/20&-191/5\\
56/5&-2/5&-267/20&136/5&-61/16\\
-15/4&22&-191/5&-61/16&163/5
\end{pmatrix}
\end{equation}
\begin{equation}\label{eq:block-d1-matrix}
K_{D,1}=\begin{pmatrix}677/24&119/240&1576/3\\119/240&18&-11647/480\\1576/3&-11647/480&61325/6\end{pmatrix}.
\end{equation}

\paragraph{Derivation of the block coefficients.}
The decompositions of $A$ and $B$ follow from the entry calculations given above. We therefore detail only the additional calculations for $D$. For $D$, the first row of $K_{D,1}$ follows from
\begin{equation}\label{eq:block-d1-check}
\begin{aligned}
 26-(-53/24)&=677/24,\\
 -37/80-(-23/24)&=119/240,\\
 2(15569/120-(-5317/40))&=1576/3.
\end{aligned}
\end{equation}
The row-sum-zero (harmonic) coefficient is
$51129/20-2(59/80)-76619/30=121/120$.
These calculations explain how the three slack blocks
$A,B,D$ reduce to the five displayed small matrices
and positive multiples of squared norms.
The dimensions are $5+3\cdot3+2=16$, so the decomposition covers all coordinates.

\subsection{Leading principal minors}
It remains to prove that the five coefficient matrices appearing above are positive definite.

For a symmetric matrix $K$, let $\Delta_j(K)$ be the determinant of its leading $j\times j$ principal block. The exact values listed in Table~\ref{tab:minors} are all positive:
\begin{longtable}{rrl}
\caption{Leading principal minors.}\label{tab:minors}\\
\toprule
$K$&$j$&$\Delta_j(K)$\\\midrule
\endfirsthead
\toprule
$K$&$j$&$\Delta_j(K)$\\\midrule\endhead
$K_A$&1&$883/96$\\ $K_A$&2&$42391/600$\\
$K_{B,0}$&1&$33/10$\\ $K_{B,0}$&2&$2969/100$\\ $K_{B,0}$&3&$1751083/2000$\\
$K_{B,1}$&1&$14/5$\\ $K_{B,1}$&2&$1748719/160000$\\
$K_{D,0}$&1&$845/96$\\ $K_{D,0}$&2&$41973071/230400$\\
$K_{D,0}$&3&$9849241061/2304000$\\ $K_{D,0}$&4&$2904628609513/92160000$\\
$K_{D,0}$&5&$3302635567136267/14745600000$\\
$K_{D,1}$&1&$677/24$\\ $K_{D,1}$&2&$29232239/57600$\\ $K_{D,1}$&3&$1052362152331/5529600$\\\bottomrule
\end{longtable}
Sylvester's criterion proves that all five small matrices are positive definite. Together with the positive coefficients in \eqref{eq:A-form}, \eqref{eq:B-form}, and \eqref{eq:D-form}, this proves strict positivity of $A,B,D$, and hence of all eighteen slack matrices.

\subsection{Exact verification procedure}\label{app:protocol}

The reported verification environment used Python~3.12.14
and SymPy~1.14.0.
The script reconstructs $H_0,H_1,H_2,C$ from the orbit tables,
forms $\widehat Q=F_3^{\mathsf T}CF_3$ by direct matrix
multiplication, and constructs all eighteen slack matrices.
Their positive definiteness is checked by exact
$LDL^{\mathsf T}$ decomposition with rational pivots.
The block identities and the fifteen displayed leading
principal minors are checked separately.

The script also verifies permutation congruence for all five
nonzero query indices, the Fourier basis sizes $1,6,16,26$,
every required Fourier sum against the generating-function
formula, the polynomial and SOS identities on all inputs,
and the certificate norm used in the error bound.
All checks use integer or rational arithmetic.
The complete implementation is given in Section~\ref{app:code}.

\subsection{Complete Python source}\label{app:code}
The following listing contains the complete exact-arithmetic verification script embedded in this manuscript. It was tested with Python~3.12.14 and SymPy~1.14.0; no floating-point tolerance is used.

\begin{lstlisting}[caption={Embedded exact rational audit source},label={lst:verify-exact5}]
"""Exact rational audit of the certificate for EXACT^5_{3,4}.

Embedded source for the exact rational audit; shown here for reproducibility.
Tested with Python 3.12.14 and SymPy 1.14.0.
No floating point tolerance is used. All 18 slack matrices are checked directly.
"""
from itertools import combinations
from collections import defaultdict
import sympy as s

R = s.Rational
sets = [[m for m in range(32) if m.bit_count() <= t] for t in range(4)]
sets = [sorted(v, key=lambda m: (m.bit_count(), m)) for v in sets]
q = [0, 1, 2, 4, 8, 16]
chi = lambda x, a: (-1)**((int(x) & int(a)).bit_count())
F = [s.Matrix(32, len(v), lambda x, j: chi(x, v[j])) for v in sets]
assert tuple(len(v) for v in sets) == (1, 6, 16, 26)

h1 = {(0,0,0):-39160,(0,1,0):1526355,(1,1,0):-1313423,(1,1,1):-23040}
h2 = {(0,0,0):-58360,(0,1,0):1526355,(0,2,0):-1298831,
      (1,1,0):-1313423,(1,1,1):-42240,(1,2,0):1021083,
      (1,2,1):1526355,(2,2,0):-2518478,(2,2,1):-1320335,(2,2,2):-33792}
ct_rows = [((0,3,0),21447408),((0,4,0),-31637000),((1,3,0),-11962875),
           ((1,3,1),-16148508),((1,4,0),16286060),((1,4,1),24857047),
           ((2,3,0),5451300),((2,3,1),6655097),((2,3,2),7901158),
           ((2,4,1),-12204683),((2,4,2),-15383580),((3,5,3),10457368)]
assert len(ct_rows) == 12 and len({key for key, _ in ct_rows}) == len(ct_rows)
ct = dict(ct_rows)

def orbit(a, b):
    a, b = int(a), int(b)
    return (*sorted((a.bit_count(), b.bit_count())), (a & b).bit_count())

H = [s.Matrix([[-1]])] + [
    s.Matrix(len(sets[t]), len(sets[t]),
             lambda j, k: R(h[orbit(sets[t][j], sets[t][k])], 19200))
    for t, h in [(1, h1), (2, h2)]
]
C = s.Matrix(32, 32, lambda x, y: R(ct.get(orbit(x, y), 0), 307200))
f = lambda x: int(x.bit_count() in (3, 4))
assert C == C.T
assert all(C[x, y] == 0 for x in range(32) for y in range(32)
           if f(x) == f(y))

T = [[s.Matrix(len(sets[t+1]), len(sets[t]),
               lambda r, c: int(sets[t+1][r] == (sets[t][c] ^ qi)))
      for qi in q] for t in (0, 1)]
Z = [[H[t] - T[t][i].T * H[t+1] * T[t][i] for i in range(6)]
     for t in (0, 1)]
Z.append([])
for qi in q:
    Dchi = s.diag(*[chi(x, qi) for x in range(32)])
    Z[2].append(H[2] + F[2].T * Dchi * C * Dchi * F[2])

for t in range(3):
    assert F[t].T * F[t] == 32 * s.eye(len(sets[t]))
    for i in range(6):
        M = Z[t][i]
        assert M == M.T
        _, piv = M.LDLdecomposition(hermitian=False)
        assert all(piv[j, j] > 0 for j in range(M.rows))
print('PASS: all 18 full slack matrices are positive definite over Q.')

assert Z[0][0] == s.Matrix([[R(499, 480)]])
assert all(Z[0][i] == s.Matrix([[R(1, 5)]]) for i in range(1, 6))
assert Z[1][0] == s.eye(6)
for t in range(3):
    for i in range(1, 6):
        def swap(a):
            return a ^ (1 | (1 << (i - 1))) if ((a & 1) > 0) != ((a & (1 << (i - 1))) > 0) else a
        perm = [sets[t].index(swap(a)) for a in sets[t]]
        assert Z[t][i] == Z[t][1].extract(perm, perm)
print('PASS: scalar identities and all certificate permutation identities.')

A, B, D = 5 * Z[1][1], Z[2][0], 5 * Z[2][1]
bt = {(0,0,0):R(33,10),(0,1,0):-R(1,50),(0,2,0):-R(181,200),
      (1,1,1):R(13,5),(1,1,0):-R(1,5),(1,2,1):-R(37,400),
      (1,2,0):-R(23,120),(2,2,2):R(141,25),(2,2,1):-R(481,600),(2,2,0):R(529,300)}
assert B == s.Matrix(16, 16,
    lambda j, k: bt[orbit(sets[2][j], sets[2][k])])

dt = {('alpha','alpha'):R(845,96),('alpha','beta'):-R(1,10),('beta','beta'):R(1987,96),
      ('alpha','u'):-R(37,80),('alpha','v'):R(14,5),('alpha','w'):-R(5,8),
      ('beta','u'):-R(333,40),('beta','v'):-R(1,10),('beta','w'):R(11,3),
      ('u','u',1):s.Integer(26),('u','u',0):-R(53,24),
      ('u','v',1):-R(37,80),('u','v',0):-R(23,24),
      ('u','w',1):R(15569,120),('u','w',0):-R(5317,40),
      ('v','v',1):R(76,5),('v','v',0):-R(14,5),
      ('v','w',1):-R(249,40),('v','w',0):R(5671,960),
      ('w','w',2):R(51129,20),('w','w',1):R(59,80),('w','w',0):-R(76619,30)}

def dtype(mask):
    return 'alpha' if mask == 0 else 'beta' if mask == 1 else 'u' if mask.bit_count() == 1 else 'v' if mask & 1 else 'w'

for j, mask_a in enumerate(sets[2]):
    for k, mask_b in enumerate(sets[2]):
        a, b = mask_a, mask_b
        ta, tb = dtype(a), dtype(b)
        if ta > tb:
            a, b, ta, tb = b, a, tb, ta
        key = (ta, tb) if ta in ('alpha', 'beta') else (ta, tb, (a & b & 30).bit_count())
        assert D[j, k] == dt[key]
print('PASS: all entries in the printed B and D orbit tables.')

KA = s.Matrix([[R(883,96), -R(76,5)], [-R(76,5), R(164,5)]])
KB0 = s.Matrix([[R(33,10), -R(1,10), -R(181,20)],
                [-R(1,10), 9, -R(38,5)],
                [-R(181,20), -R(38,5), R(306,5)]])
KB1 = s.Matrix([[R(14,5), R(119,400)], [R(119,400), R(787,200)]])
KD0 = s.Matrix([[R(845,96), -R(1,10), -R(37,20), R(56,5), -R(15,4)],
                [-R(1,10), R(1987,96), -R(333,10), -R(2,5), 22],
                [-R(37,20), -R(333,10), R(155,2), -R(267,20), -R(191,5)],
                [R(56,5), -R(2,5), -R(267,20), R(136,5), -R(61,16)],
                [-R(15,4), 22, -R(191,5), -R(61,16), R(163,5)]])
KD1 = s.Matrix([[R(677,24), R(119,240), R(1576,3)],
                [R(119,240), 18, -R(11647,480)],
                [R(1576,3), -R(11647,480), R(61325,6)]])
blocks = {'KA': KA, 'KB0': KB0, 'KB1': KB1, 'KD0': KD0, 'KD1': KD1}
expected = {
    'KA':[R(883,96),R(42391,600)],
    'KB0':[R(33,10),R(2969,100),R(1751083,2000)],
    'KB1':[R(14,5),R(1748719,160000)],
    'KD0':[R(845,96),R(41973071,230400),R(9849241061,2304000),R(2904628609513,92160000),R(3302635567136267,14745600000)],
    'KD1':[R(677,24),R(29232239,57600),R(1052362152331,5529600)]}
for name, M in blocks.items():
    minors = [M[:j, :j].det() for j in range(1, M.rows + 1)]
    assert minors == expected[name]
    assert all(v > 0 for v in minors)
print('PASS: all 15 printed leading principal minors.')


assert 3 + 2*4 + 5 == 16
assert 5 + 3*3 + 2 == 16

def split_pairs(vals, vertices):
    m = len(vertices)
    pairs = list(combinations(vertices, 2))
    rows = {j: sum(vals[e] for e in pairs if j in e) for j in vertices}
    avg = sum(rows.values()) / m
    gamma = avg / (m - 1)
    b = {j: (rows[j] - avg) / (m - 2) for j in vertices}
    h = {e: vals[e] - gamma - b[e[0]] - b[e[1]] for e in pairs}
    return gamma, b, h

def quad(K, vs):
    v = s.Matrix(vs)
    return (v.T * K * v)[0]

za = s.symbols('a0:6')
zz = s.symbols('z0:16')
alpha, beta, *us = za
gamma = sum(us) / 4
rhs = R(77,96) * alpha**2 + quad(KA, [beta, gamma]) + sum((u - gamma)**2 for u in us)
assert s.expand(quad(A, za) - rhs) == 0

values = dict(zip(sets[2], zz))
alpha = values[0]
verts = list(range(5))
us = {j: values[1 << j] for j in verts}
beta = sum(us.values()) / 5
aa = {j: us[j] - beta for j in verts}
vv = {(j, k): values[(1 << j) | (1 << k)] for j, k in combinations(verts, 2)}
gamma, bb, hh = split_pairs(vv, verts)
rhs = quad(KB0, [alpha, beta, gamma])
rhs += sum(quad(KB1, [aa[j], bb[j]]) for j in verts)
rhs += R(1351,150) * sum(v*v for v in hh.values())
assert s.expand(quad(B, zz) - rhs) == 0

J = list(range(1, 5))
beta = values[1]
us = {j: values[1 << j] for j in J}
vs = {j: values[1 | (1 << j)] for j in J}
gamma = sum(us.values()) / 4
delta = sum(vs.values()) / 4
aa = {j: us[j] - gamma for j in J}
bb = {j: vs[j] - delta for j in J}
ww = {(j, k): values[(1 << j) | (1 << k)] for j, k in combinations(J, 2)}
eta, cc, hh = split_pairs(ww, J)
rhs = quad(KD0, [alpha, beta, gamma, delta, eta])
rhs += sum(quad(KD1, [aa[j], bb[j], cc[j]]) for j in J)
rhs += R(121,120) * sum(v*v for v in hh.values())
assert s.expand(quad(D, zz) - rhs) == 0
print('PASS: A, B and D quadratic-form identities for unrestricted symbolic vectors.')


def coefficients(S, Rmask):
    poly = {(0, 0, 0): 1}
    for j in range(5):
        u, v = chi(1 << j, S), chi(1 << j, Rmask)
        new = defaultdict(int)
        for key, value in poly.items():
            for shift, sign in [((0,0,0),1), ((1,0,0),u), ((0,1,0),v), ((1,1,1),u*v)]:
                new[tuple(a+b for a, b in zip(key, shift))] += value * sign
        poly = new
    return poly

Q = F[3].T * C * F[3]
for j, S in enumerate(sets[3]):
    for k, Rmask in enumerate(sets[3]):
        poly = coefficients(S, Rmask)
        value = sum(c * (poly.get((a,b,r),0) + poly.get((b,a,r),0)) for (a,b,r), c in ct.items())
        assert R(value, 307200) == Q[j, k]
assert Q[0, 0] == R(3043, 480)
assert Q[sets[3].index(1), sets[3].index(1)] == R(24, 5)
print('PASS: all 26 x 26 required Fourier sums, without normalization ambiguity.')

pairs = list(combinations(range(5), 2))
M = s.Matrix(10, 10, lambda j, k: 1 if j == k else -R(1,4) if len(set(pairs[j]) & set(pairs[k])) == 1 else R(1,6))
assert M.eigenvals() == {s.Integer(0):1, R(5,12):4, R(5,3):5}
for x in range(32):
    w = x.bit_count()
    assert s.binomial(w,3) - 3*s.binomial(w,4) + 5*s.binomial(w,5) == f(x)
    ys = {e: s.Integer(int((x & (31 ^ ((1 << e[0]) | (1 << e[1])))) == (31 ^ ((1 << e[0]) | (1 << e[1]))))) for e in pairs}
    _, b, h = split_pairs(ys, list(range(5)))
    sos = R(5,4) * sum(v*v for v in b.values()) + R(5,3) * sum(v*v for v in h.values())
    assert sos == f(x) == quad(M, [ys[e] for e in pairs])
print('PASS: degree-five polynomial and explicit degree-three SOS on all 32 inputs.')

L = sum(abs(v) for v in C)
assert L == R(6616607, 320)
assert R(1, 4) / L**2 == R(25600, 43779488192449)
print('Entrywise absolute sum L =', L)
print('Every <=3-query algorithm has worst-case error >=', 1/(4*L**2))
print('ALL EXACT CHECKS PASSED. This does not certify historical novelty.')
\end{lstlisting}
\end{appendices}

\section*{Statements and Declarations}

\textbf{Author Contributions.} Conceptualization, G.X.; methodology, G.X., H.Z. and M.W.; validation, M.W., S.X. and M.L.; formal analysis, G.X.; investigation, G.X., H.Z. and M.W.; writing—original draft preparation, G.X., H.Z., M.W. and S.X.; writing—review and editing, G.X., H.Z., M.W. and M.L.; visualization, G.X., H.Z. and M.W.; supervision, G.X.; funding acquisition, G.X. All authors have read and agreed to the published version of the manuscript.

\textbf{Funding.} This research was funded by the National Natural Science Foundation of China, grant
numbers 62272208 and 62572221.

\textbf{Competing interests.} The authors declare no competing interests.

\textbf{Data and code availability.} No external dataset was used. The complete rational certificate, all matrix data, and the exact-arithmetic verification code supporting the results are included in the manuscript and in Appendix~\ref{app:code}. The embedded source code uses integer and rational arithmetic and is included in the submitted \TeX{} source. The reported audit was run with Python~3.12.14 and SymPy~1.14.0.

\textbf{Use of AI tools:} During manuscript preparation, the authors used OpenAI ChatGPT to assist with language editing, checking, reformulation of some derivations, and the search and computation of some numerical values. All AI-assisted mathematical content was independently verified and revised by the authors, who take full responsibility for the results and the final manuscript.

\textbf{Ethics approval and consent to participate.} Not applicable.

\end{document}